\documentclass[conference]{IEEEtran}
\usepackage{amssymb,amsmath,epsfig,graphicx,theorem}
\usepackage{amsfonts}
\usepackage{bm}
\newtheorem{Theo}{Theorem}
\newtheorem{Lem}{Lemma}

\begin{document}
\IEEEoverridecommandlockouts
\title{An Outer Bound for the Multi-Terminal Rate-Distortion Region}

\author{\authorblockN{Wei Kang \qquad \qquad Sennur Ulukus}
\authorblockA{Department of Electrical and Computer Engineering\\
University of Maryland, College Park, MD 20742\\
\emph{wkang@eng.umd.edu \qquad \qquad ulukus@umd.edu}}
\thanks{This work was supported by NSF Grants CCR $03$-$11311$, CCF $04$-$47613$ and CCF $05$-$14846$;
and ARL/CTA Grant DAAD $19$-$01$-$2$-$0011$.}}


\maketitle

\begin{abstract}
The multi-terminal rate-distortion problem has been studied  extensively. Notably, among these, Tung and Housewright have provided the best known inner and outer bounds
for the rate region under certain distortion constraints.
In this paper, we first propose an outer bound for the rate region,
and show that it is tighter than the outer bound of Tung and Housewright.
Our outer bound involves some $n$-letter Markov chain constraints,
which  cause computational difficulties.
We utilize a necessary condition for the
Markov chain constraints to obtain another outer bound, which is represented in terms of  some single-letter 
mutual information expressions evaluated over probability distributions that satisfy some single-letter conditions.
\end{abstract}

\section{The Multi-Terminal Rate-Distortion Problem}
Ever since the milestone paper of Wyner and Ziv \cite{Wyner:1976} on the rate-distortion function of a single source with  side information  at the decoder, there has been a significant amount of efforts directed towards solving a generalization of this problem, the so called multi-terminal rate-distortion problem. Despite these efforts, the problem remains open to this day.  Among all the attempts on this difficult problem,  the works by Tung \cite{Tung:1978} and Housewright \cite{Housewright:1977} (see also \cite{Berger:1978}) provide   the best inner and outer bounds so far for the rate-distortion region.

The multi-terminal rate-distortion problem can be formulated as follows. Consider a pair of discrete memoryless sources $(U, V)$, with joint distribution $p(u,v)$ defined on the finite alphabet $\mathcal{U}\times\mathcal{V}$. 
The reconstruction of the sources are built on another finite alphabet $\hat{\mathcal{U}}\times \hat{\mathcal{V}}$.
The distortion measures are defined as $d_1: \mathcal{U}\times\hat{\mathcal{U}}\longmapsto \mathbb{R}^+\cup\{0\}$ and $d_2: \mathcal{V}\times\hat{\mathcal{V}}\longmapsto \mathbb{R}^+\cup\{0\}$. 
Assume that  two distributed encoders are functions $f_1:\mathcal{U}^n\longmapsto \{1,2,\dots,M_1\}$ and 
$f_2:\mathcal{V}^n\longmapsto \{1,2,\dots,M_2\}$ and a joint decoder is the function $g:\{1,2,\dots,M_1\}
\times \{1,2,\dots,M_2\}\longmapsto\hat{\mathcal{U}}^n\hat{\times\mathcal{V}^n}$, 
where $n$ is a positive integer.
A pair of distortion levels $\mathbf{D}\triangleq(D_1,D_2)$ is said to be $\mathbf{R}$-attainable, for some rate pair
$\mathbf{R}\triangleq(R_1, R_2)$,
if for all $\epsilon>0$ and $\delta>0$, there exist, some positive integer $n$ and a set of distributed encoders and  joint decoder $(f_1, f_2, g)$ with rates $(\frac{1}{n}\log_2 M_1, \frac{1}{n}\log_2 M_2)=(R_1+\delta, R_2+\delta)$,
such that the distortion between the sources $(U^n, V^n)$ and the decoder output $(\hat{U}^n, \hat{V}^n)$
satisfies
 $\big(Ed_1(U^n, \hat{V}^n), Ed_2(V^n, \hat{V}^n)\big)<(D_1+\epsilon, D_2+\epsilon)$\footnote[1]{By $(A, B)<(C,D)$, we mean  both $A<B$ and $C<D$, and $(A, B)\le(C,D)$ is defined in the similar manner.}
where $d_1(U^n, \hat{U}^n)\triangleq\frac{1}{n}\sum_{i=1}^nd_1(U_i,\hat{U}_i)$ and 
$d_2(V^n, \hat{V}^n)\triangleq\frac{1}{n}\sum_{i=1}^nd_2(V_i,\hat{V}_i)$.
The problem here is to determine,  for a fixed $\mathbf{D}$, the set $\mathcal{R}(\mathbf{D})$ of all
rate pairs
$\mathbf{R}$, for which $\mathbf{D}$ is $\mathbf{R}$-attainable.

We restate the outer bound provided in \cite{Tung:1978} and \cite{Housewright:1977}  in the following theorem.
\begin{Theo}[\cite{Tung:1978,Housewright:1977}]\label{btb}
$\mathcal{R}(\mathbf{D})\subseteq\mathcal{R}_{\mathrm{out,1}}(\mathbf{D})$, where $\mathcal{R}_{\mathrm{out,1}}(\mathbf{D})$
is the set of all $\mathbf{R}$ such that there exists a pair  of discrete random variables $(X_1,X_2)$,  for which 
the following three conditions are satisfied:
\begin{enumerate}
\item The joint distribution satisfies
\begin{align}
X_1\rightarrow U&\rightarrow V\\
U&\rightarrow V\rightarrow X_2
\end{align}
\item  The rate pair satisfies
\begin{align}
R_1&\ge I(U,V; X_1|X_2)\\
R_2&\ge I(U,V; X_2|X_1)\\
R_1+R_2&\ge I(U,V; X_1,X_2)
\end{align}
\item There exists $\big(\hat{U}(X_1,X_2), \hat{V}(X_1,X_2)\big)$ such that 
$\big(Ed_1(U, \hat{U}), Ed_2(V, \hat{V}\big))\le \mathbf{D}$.
\end{enumerate}
\end{Theo}
An  inner bound is also given in \cite{Tung:1978} and \cite{Housewright:1977} as follows.
\begin{Theo}[\cite{Tung:1978,Housewright:1977}]
$\mathcal{R}(\mathbf{D})\supseteq\mathcal{R}_{\mathrm{in}}(\mathbf{D})$, where $\mathcal{R}_{\mathrm{in}}(\mathbf{D})$
is the set of all $\mathbf{R}$ such that there exists a pair  of discrete random variables $(X_1,X_2)$,  for which 
the following three conditions are satisfied:
\begin{enumerate}
\item The joint distribution satisfies
\begin{align}
X_1\rightarrow U&\rightarrow V\rightarrow X_2
\end{align}
\item  The rate pair satisfies
\begin{align}
R_1&\ge I(U,V; X_1|X_2)\\
R_2&\ge I(U,V; X_2|X_1)\\
R_1+R_2&\ge I(U,V; X_1,X_2)
\end{align}
\item There exists $\big(\hat{U}(X_1,X_2), \hat{V}(X_1,X_2)\big)$ such that 
$\big(Ed_1(U, \hat{U}), Ed_2(V, \hat{V})\big)\le \mathbf{D}$.
\end{enumerate}
\end{Theo}

We note that the inner and outer bounds agree on both the second condition, i.e.,~the rate constraints in terms of some mutual information expressions, and the third condition, i.e.,~ the reconstruction functions.  However, the first condition in these two bounds constraining the underlying probability distributions $p (x_1, x_2|u, v)$ are different. It is easy to see that the Markov chain condition in the inner bound, i.e.,~$X_1\rightarrow U\rightarrow V\rightarrow X_2$,
implies the Markov chain conditions in the outer bound, i.e.,~$X_1\rightarrow U\rightarrow V$ and $U\rightarrow V\rightarrow X_2$.  Hence, if we define
\begin{align}
\mathcal{S}_{out,1}&\triangleq\{p(x_1,x_2|u,v): X_1\rightarrow U\rightarrow V\text{ and }U\rightarrow V\rightarrow X_2\}\\
 \mathcal{S}_{in}&\triangleq\{p(x_1,x_2|u,v): X_1\rightarrow U\rightarrow V\rightarrow X_2\}
\end{align}
then, 
\begin{equation}
\mathcal{S}_{out,1}\supseteq\mathcal{S}_{in}
\end{equation}

Using the time-sharing argument, a convexification of the inner bound $\mathcal{R}_{in}(\mathbf{D})$ yields another inner bound  $\mathcal{R}_{in}'(\mathbf{D})$, which is larger. This new inner bound may be expressed as a function of $\mathcal{S}_{in}$ and $\mathbf{D}$ as follows,
\begin{equation}
\mathcal{R}(\mathbf{D})\supseteq\mathcal{R}'_{in}(\mathbf{D})=\mathcal{F}(\mathcal{S}_{in},\mathbf{D}) \supseteq\mathcal{R}_{in}(\mathbf{D})
\end{equation}
where, using a time sharing random variable $Q$, which is 
 known by encoders and decoder,  $\mathcal{F}(\mathcal{S}_{in},\mathbf{D})$ is defined as,
\begin{align}
\mathcal{F}(\mathcal{S}_{in},\mathbf{D})\triangleq& \bigcup_{\mathbf{p}\in\mathcal{P}(\mathcal{S}_{in},\mathbf{D})}\mathcal{C}(\mathbf{p})\\
\mathbf{p}\triangleq&p(x_1,x_2,q|u,v)=p_q(x_1,x_2|u,v)p(q)\\
\mathcal{P}(\mathcal{S}_{in},\mathbf{D})\triangleq&\left\{\mathbf{p}:\begin{array}{l}
p_q(x_1,x_2|u,v)\in\mathcal{S}_{in};\\
 \exists \big(\hat{U}(X_1,X_2, Q), \hat{V}(X_1,X_2, Q)\big), \\
 \text{ s.t. }\big(Ed_1(U, \hat{U}), Ed_2(V, \hat{V})\big)\le \mathbf{D}
\end{array}\right\}\\
\mathcal{C}(\mathbf{p})\triangleq&\left\{(R_1, R_2):\!\!\!\!\begin{array}{rcl}R_1\ge I(U,V;X_1|X_2, Q)\\R_2\ge I(U,V;X_2|X_1, Q)\\R_1+R_2\ge I(U,V;X_1,X_2|Q)\end{array}\right\}
\end{align}

 In \cite{Housewright:1977}, it was shown that $\mathcal{R}_{out,1}(\mathbf{D})$ is convex, which means that $\mathcal{R}_{out,1}(\mathbf{D})$  can also be represented as a function of  $ \mathcal{S}_{out,1}$ and $\mathbf{D}$ as follows.
 \begin{equation}
\mathcal{R}_{out,1}(\mathbf{D})=\mathcal{F}(\mathcal{S}_{out,1},\mathbf{D}) 
 \end{equation}
 
 Therefore, we conclude that the gap between the inner and the outer bounds comes only from the difference between the feasible sets of the probability distributions $p(x_1,x_2|u,v)$. In the next section, we will provide a tighter outer bound for the rate region in the sense that it can be represented using the same mutual information expressions, however, on a smaller feasible set for $p(x_1,x_2|u,v)$ than $\mathcal{R}_{out,1}(\mathbf{D})$.

\section{A New Outer Bound}
We propose a new outer bound as follows.
\begin{Theo}\label{nletter2}
$\mathcal{R}(\mathbf{D})\subseteq\mathcal{R}_{\mathrm{out,2}}(\mathbf{D})$, where $\mathcal{R}_{\mathrm{out,2}}(\mathbf{D})$
is the set of all $\mathbf{R}$ such that there exist some positive integer $n$,
 and discrete random variables $Q, X_1, X_2$ for which 
the following three conditions are satisfied:
\begin{enumerate}
\item The joint distribution satisfies
\begin{align}
p(u^n,& v^n, x_1, x_2,  q)\nonumber\\
=&p(q)p(x_1|u^n, q)p(x_2|v^n, q)\prod_{i=1}^n p(u_i, v_i)\label{dist}
\end{align}
\item The rate pair satisfies
\begin{align}
R_1&\ge I(U_1,V_1; X_1|X_2, Q)\label{r1}\\
R_2&\ge I(U_1,V_1; X_2|X_1, Q)\label{r2}\\
R_1+R_2&\ge I(U_1,V_1; X_1,X_2|Q)\label{r1r2}
\end{align}
where $(U_1, V_1)$ is the first sample of the $n$-sequences $(U^n, V^n)$.
\item There exists $\big(\hat{U}(X_1,X_2, Q), \hat{V}(X_1,X_2, Q)\big)$ such that $\big(Ed_1(U, \hat{U}), Ed_2(V, \hat{V})\big)\le \mathbf{D}$.
\end{enumerate}
or equivalently,
\begin{equation}
\mathcal{R}_{out,2}(\mathbf{D})=\mathcal{F}(\mathcal{S}_{out,2},\mathbf{D}) 
\end{equation}
where
\begin{equation}
\mathcal{S}_{out,2}\triangleq \{p(x_1,x_2|u_1,v_1): X_1\rightarrow U^n\rightarrow V^n\rightarrow X_2\}
\end{equation}
\end{Theo}
\begin{proof}
Consider an arbitrary set of distributed encoders and joint decoder $(f_1, f_2, g)$ 
with reconstructions $(\hat{U}^n, \hat{V}^n)=g(W, Z)$, where $W=f_1(U^n)$ and $Z=f_2(V^n)$,  such that  $\big(Ed_1(U^n, \hat{V}^n), Ed_2(V^n, \hat{V}^n)\big)<(D_1+\epsilon, D_2+\epsilon)$.  Here, we use $R_1=\frac{1}{n}\log_2(M_1)=\frac{1}{n}\log_2(|W|)$ and $R_2=\frac{1}{n}\log_2(M_2)=\frac{1}{n}\log_2(|Z|)$.

We define the auxiliary random variables $X_{1i}=(W, U^{i-1})$ and $X_{2i}=(Z, V^{i-1})$. Then, we have
\begin{align}
\log_2&(M_1)\ge H(W)\nonumber\\&=I(U^n,V^n; W)\nonumber\\
&\overset{1)}{\ge} I(U^n,V^n; W|Z)\nonumber\\
&=\sum_{i=1}^n I(U_i, V_i; W|Z, U^{i-1}, V^{i-1})\nonumber\\
&= \sum_{i=1}^n I(U_i, V_i; W,Z| U^{i-1}, V^{i-1}) \nonumber\\
&\quad
-I(U_i, V_i; Z| U^{i-1}, V^{i-1})\nonumber\\
&\overset{2)}{=} \sum_{i=1}^n I(U_i, V_i; W,Z| U^{i-1}, V^{i-1})
-I(U_i, V_i; Z| V^{i-1})\nonumber\\
&= \sum_{i=1}^n I(U_i, V_i; W,Z, U^{i-1}|  V^{i-1})
\nonumber\\&\quad
-I(U_i, V_i; U^{i-1}|V^{i-1})-I(U_i, V_i; Z| V^{i-1})\nonumber\\
&\overset{3)}{=}\sum_{i=1}^n I(U_i, V_i; W,Z, U^{i-1}|  V^{i-1})
-I(U_i, V_i; Z| V^{i-1})\nonumber\\
&=\sum_{i=1}^n I(U_i, V_i; W, U^{i-1}| Z,V^{i-1})\nonumber\\
&=\sum_{i=1}^n I(U_i, V_i;X_{1i}|X_{2i})
\end{align}
where
\begin{enumerate}
\item follows from the fact that 
$W\rightarrow U^n\rightarrow V^n\rightarrow Z$. 
We observe that the equality holds
 when $W$ is independent of $Z$;
\item from the fact that 
\begin{align}
p(z|u_i,v_i,v^{i-1})&=p(z|u_i,v_i,u^{i-1},v^{i-1})
\end{align}
\item from the memoryless property of the sources.
\end{enumerate}

Using a symmetrical argument, we obtain
\begin{equation}
\log_2(M_2)\ge \sum_{i=1}^nI(U_i,V_i; X_{2i}|X_{1i})
\end{equation}

Moreover,
\begin{align}
\log_2&(M_1M_2)\ge H(W,Z)\nonumber\\=&I(U^n,V^n; W,Z)\nonumber\\
=&\sum_{i=1}^nH(U_i,V_i)-H(U_i,V_i|W,Z,U^{i-1},V^{i-1})\nonumber\\=&\sum_{i=1}^n
I(U_i,V_i; X_{1i},X_{2i})
\end{align}

We introduce a time-sharing random variable $Q$, 
which is uniformly distributed on $\{1,\dots, n\}$ and independent of $U^n$ and $V^n$. 
Let the random variables $X_1$ and $X_2$ be such that
\begin{equation}
p(x_{1i},x_{2i}|u_i, v_i, u^{-i}, v^{-i})=p(x_1, x_2|u_1, v_1, u^{-1}, v^{-1}, Q=i)
 \end{equation}
 where $U^{-i}\triangleq \{U_1,\dots,U_{i-1}, U_{i+1},\dots, U_n\}$ and $V^{-i}$ is defined similarly.
 Then,  
 \begin{align}
 \sum_{i=1}^nI(U_i,V_i; X_{1i}|X_{2i})&=nI(U_1,V_1; X_1|X_2, Q)\\
 \sum_{i=1}^nI(U_i,V_i; X_{2i}|X_{1i})&=nI(U_1,V_1; X_2|X_1, Q)\\
 \sum_{i=1}^nI(U_i,V_i; X_{1i},X_{2i})&=nI(U_1,V_1; X_1,X_2| Q)
 \end{align}
 
The reconstruction pair $(\hat{U},\hat{V})$ is  defined as follows. When $Q=i$, $(\hat{U},\hat{V})\triangleq(\hat{U}_i,\hat{V}_i)$, i.e.,~the $i$-th letter of $(\hat{U}^n, \hat{V}^n)=g(W, Z)$. $(\hat{U}_i,\hat{V}_i)$ is a function of $(W,Z)$, and,  therefore, it is  a function of $(X_1, X_2, Q)$. Hence, we have that $(\hat{U},\hat{V})$ is a function of $(X_1, X_2, Q)$, i.e.,~$\big(\hat{U}(X_1, X_2, Q),\hat{V}(X_1, X_2, Q)\big)$.
It is easy to see that  
\begin{align}
\big(Ed_1(U, \hat{U}), Ed_2(V, \hat{V})\big)&=\big(Ed_1(U^n, \hat{V}^n), Ed_2(V^n, \hat{V}^n)\big)\nonumber\\&<(D_1+\epsilon, D_2+\epsilon)
\end{align}
which completes the proof. 
\end{proof}

Next, we state and prove that our outer bound given in Theorem \ref{nletter2} is tighter than
the outer bound of \cite{Tung:1978} and \cite{Housewright:1977} given in Theorem \ref{btb}.
\begin{Theo}\label{tighter}
\begin{equation}
\mathcal{R}_{\mathrm{out,1}}(\mathbf{D})\supseteq\mathcal{R}_{\mathrm{out,2}}(\mathbf{D})
\end{equation}
\end{Theo}
\begin{proof}
Here, we provide two proofs. First, 
we prove this theorem by construction. For every $(R_1, R_2)$ point in $\mathcal{R}_{\mathrm{out,2}}(\mathbf{D})$,
there exist random variables $Q, X_1, X_2$ satisfying (\ref{dist}), $(R_1, R_2)$ pair satisfying (\ref{r1}), (\ref{r2}) and
(\ref{r1r2}), and a reconstruction pair $\big(\hat{U}(X_1,X_2, Q), \hat{V}(X_1,X_2, Q)\big)$ such that $\big(Ed_1(U, \hat{U}), Ed_2(V, \hat{V})\big)\le \mathbf{D}$. According to \cite{Housewright:1977}, let $X_1'=(X_1, Q)	$ and $X_2'=(X_2, Q)$. Then, $X_1'$ and $X_2'$ satisfy the first 
condition of Theorem \ref{btb}. Moreover,
\begin{equation}
R_1\ge I(U,V; X_1|X_2, Q)=I(U,V;X_1'|X_2')
\end{equation}
and similarly,
\begin{equation}
R_2\ge I(U,V; X_2|X_1, Q)=I(U,V;X_2'|X_1')
\end{equation}
and finally,
\begin{align}
R_1+R_2 &\ge I(U,V; X_1,X_2|Q)\nonumber\\
&=H(U, V|Q)-H(U, V|X_1, X_2, Q)\nonumber\\
&\overset{1)}{=}H(U, V)-H(U, V|X_1, X_2, Q)\nonumber\\
&= H(U, V)-H(U, V|X_1', X_2')\nonumber\\
&=I(U,V; X_1',X_2')
\end{align}
where
1) follows from the fact that $Q$ is independent of $(U, V)$.
$(\hat{U}, \hat{V})$ is a function of $(X_1, X_2, Q)$, and, therefore,  it is  a function of $(X_1', X_2')=\big((X_1,Q), ( X_2, Q)\big)$.

Hence, for every rate pair $(R_1, R_2)\in \mathcal{R}_{\mathrm{out,2}}(\mathbf{D})$,
there exist random variables $X_1', X_2'$ satisfying the first condition of Theorem \ref{btb}, and
$(R_1, R_2)$ pair satisfies the second condition of Theorem \ref{btb},  with the distortion satisfying the third condition of Theorem \ref{btb}. In other words, 
$(R_1, R_2)\in \mathcal{R}_{\mathrm{out,1}}(\mathbf{D})$, proving the theorem.

An alternative proof comes from the comparison of $\mathcal{S}_{out,1}$ and $\mathcal{S}_{out,2}$,   the feasible sets of probability distributions $p(x_1,x_2|u_1,v_1)$\footnote{In $\mathcal{S}_{out,1}$, the probability distribution is $p(x_1,x_2|u,v)$. Here, we just rename $U=U_1$ and $V=V_1$.}.  We note that $X_1\rightarrow U^n\rightarrow V^n\rightarrow X_2$ implies $X_1\rightarrow U_1\rightarrow V_1$ and $ U_1\rightarrow V_1\rightarrow X_2$, which means that 
\begin{equation}
\mathcal{S}_{out,1}\supseteq\mathcal{S}_{out,2}
\end{equation}
and therefore
\begin{equation}
\mathcal{R}_{\mathrm{out,1}}(\mathbf{D})=\mathcal{F}(\mathcal{S}_{out,1},\mathbf{D})\supseteq\mathcal{F}(\mathcal{S}_{out,2},\mathbf{D})=\mathcal{R}_{\mathrm{out,3}}(\mathbf{D})
\end{equation}
\end{proof}
\section{A Single-letter Outer Bound}

The proposed  outer bound $\mathcal{R}_{\mathrm{out,2}}(\mathbf{D})$ 
is an $n$-letter bound because the feasible set of the probability distribution,  $\mathcal{S}_{out,2}$, is characterized by an $n$-letter Markov chain
constraint $X_1\longrightarrow U^n \longrightarrow V^n \longrightarrow X_2$, which is practically incomputable when $n$ is sufficiently large. 
In the rest of this section, we will find a single-letter necessary condition for this
$n$-letter Markov chain.
By doing this, we will obtain a single-letter outer bound for $\mathcal{R}(\mathbf{D})$.

We introduce our matrix notation for probability distributions \cite{Kang:2005, Kang:2006a}.
For a pair of discrete random variables $X$ and $Y$,
which take values in $\mathcal{X}=\{x_1, x_2,\dots, x_k\}$
and $\mathcal{Y}=\{y_1, y_2,\dots, y_l\}$, respectively,
the joint distribution matrix $P_{XY}$ is defined as
$P_{XY}(i,j)\triangleq Pr(X=x_i, Y=y_j)$,
where $P_{XY}(i,j)$ denotes the $(i,j)$-th element of the matrix $P_{XY}$.
Similarly, we define $P_{XY|z}$ as $P_{XY|z}(i,j)\triangleq Pr(X=x_i, Y=y_j|Z=z)$.
The marginal distribution of a random variable $X$ is defined as
a diagonal matrix with $P_{X}(i,i)\triangleq Pr(X=x_i)$.
The vector-form marginal distribution is defined as $p_X(i)\triangleq Pr(X=x_i)$,
i.e.,~$p_X=P_X \mathbf{e}$, where $\mathbf{e}$ is a vector of all ones.
$p_X$ can also be  defined as $p_X\triangleq P_{XY}$ for some $Y$
where the size of the alphabet of $Y$, $|\mathcal{Y}|$,  is equal to one.
A column vector $p_{X|z}$ is defined as $p_{X|z}(i)\triangleq Pr(X=x_i|Z=z)$,
or equivalently, $p_{X|z}(i)\triangleq P_{XY|z}$ for some $Y$
where the size of the alphabet of $Y$, $|\mathcal{Y}|$,  is equal to one.
We define a new quantity, $\tilde{P}_{XY}$, as
\begin{equation}
\tilde{P}_{XY}=P_X^{-\frac{1}{2}}P_{XY}P_Y^{-\frac{1}{2}}\label{def}
\end{equation}
Since $p_X\triangleq P_{XY}$ for some $Y$ 
where the size of the alphabet of $Y$, $|\mathcal{Y}|$,  is equal to one,
we define
\begin{equation}
\tilde{p}_{X}=P_X^{-\frac{1}{2}}P_{XY}P_Y^{-\frac{1}{2}}=P_X^{-\frac{1}{2}}p_{X}\label{def1}
\end{equation}
The conditional distributions $\tilde{P}_{XY|z}$ and $\tilde{p}_{X|z}$ can be defined similarly.

In \cite{Kang:2005}, we provided a new data processing inequality, a necessary condition for a Markov chain,
as follows.
\begin{Theo}[\cite{Kang:2005}]\label{newdata}
If $X\rightarrow Y\rightarrow Z$, then
\begin{align}
\lambda_i(\tilde{P}_{XZ})\le\lambda_i(\tilde{P}_{XY})\lambda_2(\tilde{P}_{YZ})&\le\lambda_i(\tilde{P}_{XY})
 \end{align}
where  $i=2,\dots, \mathrm{rank}(\tilde{P}_{XZ})$,
and where $\lambda_{i}(\cdot)$ denotes the $i$-th largest singular value of a matrix.
\end{Theo}

We have also shown in \cite{Kang:2005} the following lemma.
\begin{Lem}[\cite{Kang:2005}]\label{iid}
 For a pair of i.i.d. sequences $(X^k, Y^k)$ characterized by a joint distribution $P_{XY}$,
the ordered singular values of $\tilde{P}_{X^kY^k}$ are
\begin{equation}
\{1, \lambda_2(\tilde{P}_{XY}), \dots, \lambda_2(\tilde{P}_{XY}),\dots\}\nonumber
\end{equation}
where the second through the $k+1$-st singular values are all equal to $\lambda_2(\tilde{P}_{XY})$. 
\end{Lem}

Combining Theorem \ref{newdata} and Lemma \ref{iid},  we have the following theorem.
\begin{Theo}[\cite{Kang:2006a}]\label{nec}
Let $(U^n, V^n)$ be a pair of i.i.d. sequences of length $n$, and let the random variables $X_1$ and $X_2$ satisfy
$X_1\longrightarrow U^n \longrightarrow V^n \longrightarrow X_2$.  
Then, for $i=2,\dots, \min(|\mathcal{X}_1|, |\mathcal{X}_2|)$,
\begin{align}
\lambda_{i}(\tilde{P}_{X_1X_2})&\le\lambda_2(\tilde{P}_{UV})\label{cond1}\\
\lambda_{i}(\tilde{P}_{X_1X_2|u_1})&\le\lambda_2(\tilde{P}_{UV})\label{cond2}\\
\lambda_{i}(\tilde{P}_{X_1X_2|v_1})&\le\lambda_2(\tilde{P}_{UV})\label{cond3}\\
\lambda_{i}(\tilde{P}_{X_1X_2|u_1v_1})&\le\lambda_2(\tilde{P}_{UV})\label{cond4}
\end{align}
or equivalently
\begin{equation}
\mathcal{S}_{out,2}\subseteq\mathcal{S}_{out,3}
\end{equation}
where 
\begin{align}
\mathcal{S}_{out,3}\triangleq&\{p(x_1,x_2|u_1,v_1):\nonumber\\
&\quad (\ref{cond1}), (\ref{cond2}), (\ref{cond3}), \text{ and } (\ref{cond4}) \text{ are satisfied}\}
\end{align}
\end{Theo}
\begin{proof}
The proof of (\ref{cond1}) follows from our result in \cite{Kang:2005}. 
In (\ref{cond2}), $\tilde{P}_{X_1X_2|u_1}=
\tilde{P}_{X_1U^{-1}|u_1}\tilde{P}_{U^{-1}V^n|u_1}\tilde{P}_{V^nX_2}$. 
It can be shown that $\tilde{P}_{U^{-1}V^n|u_1}=\tilde{P}_{U^{-1}V^{-1}}\otimes \tilde{p}_{V_1|u_1}^T$
where $\otimes$ represents the Kronecker product.
Since $\tilde{p}_{V_1|u_1}^T$ is a vector with singular value $1$, 
$\lambda_{i}(\tilde{P}_{U^{-1}V^n|u_1})=\lambda_{i}(\tilde{P}_{U^{-1}V^{-1}})$. 
Applying Theorem \ref{newdata} and \ref{nec}, (\ref{cond2}) is proven.
In a similar manner, (\ref{cond3}) can be shown. 
In (\ref{cond4}), $\tilde{P}_{X_1X_2|u_1v_1}=
\tilde{P}_{X_1U^{-1}|u_1v_1}\tilde{P}_{U^{-1}V^{-1}|u_1v_1}\tilde{P}_{V^{-1}X_2|u_1v_1}$.
Since $(U^n, V^n)$ is a pair of  i.i.d. sequences, $\tilde{P}_{U^{-1}V^{-1}|u_1v_1}=\tilde{P}_{U^{-1}V^{-1}}$.
Then (\ref{cond4}) is a direct result of Theorem \ref{newdata} and \ref{nec}.
\end{proof}

From the above discussion, we obtain the main result of our paper, 
which is stated in the following theorem.

\begin{figure*}\label{fig1}
\centering
\includegraphics[width=6in]{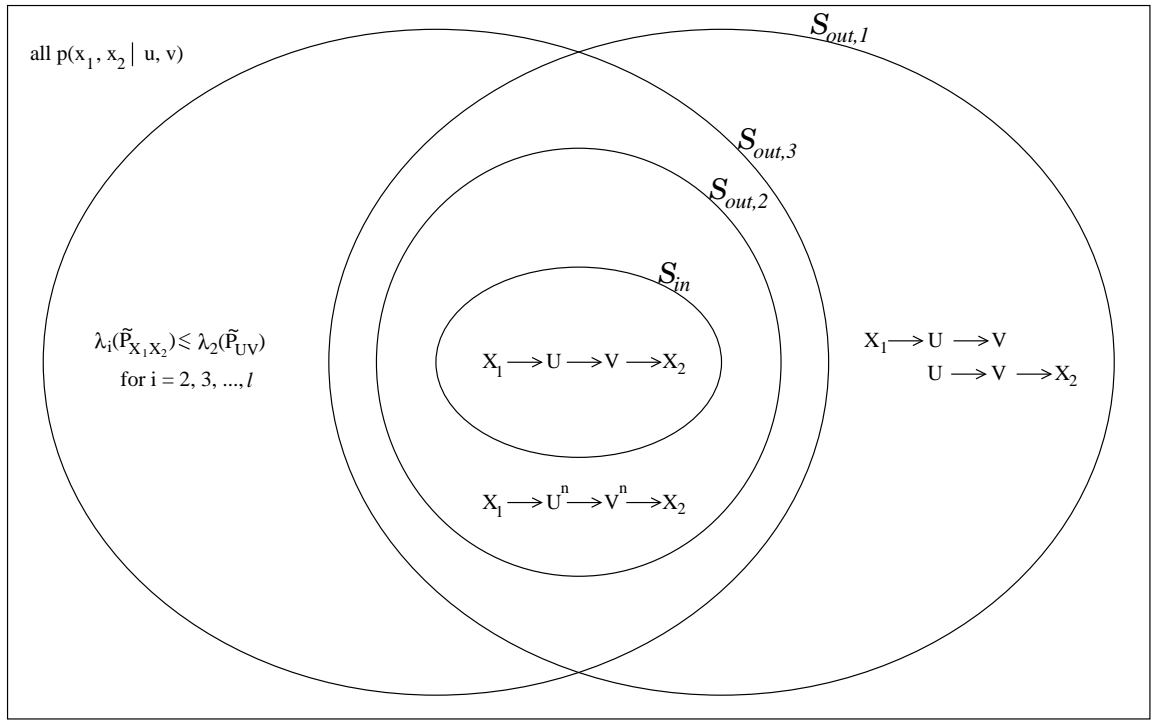}
\caption{Different sets of probability distributions $p(x_1,x_2|u,v)$.}
\end{figure*}

\begin{Theo}
$\mathcal{R}(\mathbf{D})\subseteq\mathcal{R}_{\mathrm{out,3}}(\mathbf{D})$, where $\mathcal{R}_{\mathrm{out,3}}(\mathbf{D})$
is the set of all $\mathbf{R}$ such that there exists some positive integer $n$,
 and there exist discrete random variable $Q$ independent of $(U^n, V^n)$,  and discrete random variables $X_1, X_2$ for which 
the following three conditions are satisfied:
\begin{enumerate}
\item The joint distribution satisfies,  for $i=2,\dots, \min(|\mathcal{X}_1|, |\mathcal{X}_2|)$, 
\begin{align}
\lambda_{i}(\tilde{P}_{X_1X_2|q})&\le\lambda_2(\tilde{P}_{UV})\\
\lambda_{i}(\tilde{P}_{X_1X_2|uq})&\le\lambda_2(\tilde{P}_{UV})\\
\lambda_{i}(\tilde{P}_{X_1X_2|vq})&\le\lambda_2(\tilde{P}_{UV})\\
\lambda_{i}(\tilde{P}_{X_1X_2|uvq})&\le\lambda_2(\tilde{P}_{UV})
\end{align}
\item The rate pair satisfies
\begin{align}
R_1&\ge I(U,V; X_1|X_2, Q)\\
R_2&\ge I(U,V; X_2|X_1, Q)\\
R_1+R_2&\ge I(U,V; X_1,X_2|Q)
\end{align}
\item There exists $\big(\hat{U}(X_1,X_2, Q), \hat{V}(X_1,X_2, Q)\big)$ such that $\big(Ed_1(U, \hat{U}), Ed_2(V, \hat{V})\big)\le \mathbf{D}$.
\end{enumerate}
or equivalently,
\begin{equation}
\mathcal{R}_{out,3}(\mathbf{D})=\mathcal{F}(\mathcal{S}_{out,3},\mathbf{D}) 
\end{equation}
\end{Theo}
From Theorem \ref{nec}, we have that 
\begin{equation}
\mathcal{S}_{out,2}\subseteq\mathcal{S}_{out,3}
\end{equation}
and therefore
\begin{equation}
\mathcal{R}_{\mathrm{out,2}}(\mathbf{D})=\mathcal{F}(\mathcal{S}_{out,2},\mathbf{D}) \subseteq\mathcal{R}_{\mathrm{out,3}}(\mathbf{D})=\mathcal{F}(\mathcal{S}_{out,3},\mathbf{D}) 
\end{equation}
From Theorem \ref{tighter}, we know that
\begin{equation}
\mathcal{S}_{out,2}\subseteq\mathcal{S}_{out,1}
\end{equation}
and
\begin{equation}
\mathcal{R}_{\mathrm{out,2}}(\mathbf{D})=\mathcal{F}(\mathcal{S}_{out,2},\mathbf{D}) \subseteq\mathcal{R}_{\mathrm{out,1}}(\mathbf{D})=\mathcal{F}(\mathcal{S}_{out,1},\mathbf{D}) 
\end{equation}
So far, we have not been able to determine whether
$\mathcal{S}_{out,3}\subseteq\mathcal{S}_{out,1}$ or $\mathcal{S}_{out,1}\subseteq\mathcal{S}_{out,3}$, 
however, we know that there exists some probability distribution $p(x_1,x_2|u,v)$, which belongs to $\mathcal{S}_{out,1}$, but does not belong to $\mathcal{S}_{out,3}$. For example, assume $\lambda_2(\tilde{P}_{UV})<1$. Let $X_1=(f_1(U), S)$ and $X_2=(f_2(V), S)$. We note that $(X_1, X_2, U,V)$ satisfy $X_1\rightarrow U\rightarrow V$ and $U\rightarrow V\rightarrow X_2$, i.e.,~ $p(x_1,x_2|u,v)\in \mathcal{S}_{out,1}$. But, $(X_1, X_2)$ contain common information $S$, which means that $\lambda_2(\tilde{P}_{X_1X_2})=1> \lambda_2(\tilde{P}_{UV})$ \cite{Witsenhausen:1975}, and therefore, $p(x_1,x_2|u,v)\notin \mathcal{S}_{out,3}$.  Based on this observation,  we can see that introducing $\mathcal{S}_{out,3}$ helps us rule out some unachievable probability distributions that may exist in $\mathcal{S}_{out,1}$. 
The relation between different feasible sets of probability distributions $p(x_1,x_2|u,v)$ is illustrated in  Figure 1.

Finally, we note that, we can obtain a tighter outer bound in terms of the function $\mathcal{F}(\cdot, \mathbf{D})$ where the set argument is the intersection of $\mathcal{S}_{out,1}$ and $\mathcal{S}_{out,3}$, i.e.,
\begin{align}
\mathcal{R}_{out, 1\cap 3}(\mathbf{D})&\triangleq\mathcal{F}(\mathcal{S}_{out,1}\cap\mathcal{S}_{out,3}, \mathbf{D})
\end{align}
It is straightforward to see that this outer bound $\mathcal{R}_{out, 1\cap 3}(\mathbf{D})$ is in general tighter than the outer bound $\mathcal{F}(\mathcal{S}_{out,1}, \mathbf{D})$  of Tung and Housewright.
\section{Conclusion}
 Tung and Housewright have provided inner and outer bounds for the multi-terminal rate-distortion region in late 1970s.
In this paper, we first proposed an outer bound for the rate region,
and showed that it is tighter than the outer bound of Tung and Housewright.
Our outer bound involves some $n$-letter Markov chain constraints,
and is  not computationally practical.
To avoid this problem,
we utilized a single-letter necessary condition for the
Markov chain to obtain another outer bound
for the rate region, which is represented in terms of  some single-letter 
mutual information expressions.

\bibliographystyle{IEEEtran}
\bibliography{IEEEabrv,ref}

\end{document}